\documentclass[twocolumn,11pt]{article}
\usepackage{graphicx}
\usepackage{amsmath}
\newcommand{\R}{{\rm I}\!{\rm R}} 

\begin{document}
\global\def\refname{{\normalsize \it References:}}
\baselineskip 12.5pt
%
%
%
\title{\LARGE \bf The Nash equilibrium of forest ecosystems}

\date{October 30, 2016}

\author{\hspace*{-10pt}
\begin{minipage}[t]{2.7in} \normalsize \baselineskip 12.5pt
\centerline{ALEXANDER\ K. GUTS} \centerline{Dostoevsky Omsk State
University} \centerline{Department of Computer Sciences}
\centerline{Chair of Cybernetics} \centerline{Mira Prospect 55-a,
644077 Omsk} \centerline{RUSSIA} \centerline{aguts@mail.ru}
\end{minipage} \kern 0in
\begin{minipage}[t]{2.7in} \normalsize \baselineskip 12.5pt
\centerline{LUDMILA\ A. VOLODCHENKOVA} \centerline{Dostoevsky Omsk
State University} \centerline{Department of Computer Sciences}
\centerline{Chair of Cybernetics} \centerline{Mira Prospect 55-a,
644077 Omsk} \centerline{RUSSIA}
\centerline{volodchenkova2007@yandex.ru}
\end{minipage}
\\ \\ \hspace*{-10pt}
\begin{minipage}[b]{5.7in} \normalsize
\baselineskip 12.5pt {\it Abstract:}
To find the possible equilibrium states of forest
 ecosystems one
are suggested to use the theory of differential games.
 At within
the 4-tier model of mosaic forest communities it
 established the
existence  of the Nash equilibrium states in such ecosystems.
\\ [4mm] {\it Key--Words:}
Forest ecosystem, the equilibrium of the ecosystem,
differential
game, Nash equilibrium
\end{minipage}
\vspace{-10pt}}

\maketitle

\thispagestyle{empty} 
%
%
\section{Introduction}
\label{S1} \vspace{-4pt}

As a rule, the {\it stationary equilibrium state} of the
 system,
or {\it stationary equilibrium}, is  {\it stationary state}
 for
which characterizing its parameter $ x (t) $ does not
 change with
time, i.~e.
$$
\frac{dx}{dt} = 0.
$$
However, the systems are often controlled by external factors $
u_1, ..., u_N, $ and in fact their dynamics is described by the
differential equation of the form
$$
\frac{dx}{dt} = f(t, x, u_1, ..., u_N).
$$
In this case, it can  consider this equation in the
 framework of
optimal control theory, and moreover, in the framework of the
theory differential games, and to find the so-called the {\it Nash
equilibrium}.

In the theory of differential games each controlling factor
 $u_i$ is considered to be in possession of the
 {\it player} $i$
  who  tries to use it to affect the system so to have
  a maximal winning or minimal losing. Player's wining/lossing
  is described some given function
$J_i(x,u_1,...,u_N)$. Clearly, in reality, it is difficult to
suggests that the factors can be changed completely
independently
from each other, and therefore, in  the system can be
  installed
in some sense of equilibrium.

Nash equilibrium  in this case means that if each
player is trying
to unilaterally change their management strategy in
 the while
other players policy remains unchanged, it  has the
 worst record
(greater loss).

Forest ecosystem dynamics can also be described by the
differential equation with external control factors. As external
controlling factors may be considered such characteristics of
forest communities as a mosaic state  $m$, interspecific and
intraspecific competition $k$, the impact of the anthropogenic $a$
and soil moisture $w$.

It is natural to try to establish the existence of Nash
equilibrium in forest ecosystems with external control factors
$k,m,a,w$.

\section{Model of 4-tier mosaic forest}
\label{S2} \vspace{-4pt}

In  \cite{vgu0,vgu1} was offered the next model 4-tier mosaic
forest communities, characterized by productivity $x$:
\begin{equation} \label{vga7}     
\frac{dx}{dt} = -\frac{\partial}{\partial x} V(x,k,m,a,w),
\end{equation}
where
$$
V(x,k,m,a,w) =
$$
\begin{equation} \label{vga8}
=\frac{\alpha}{6} x^6 + kx^4 + mx^3 + ax^2 + wx,
\end{equation}
$$
\alpha=\alpha_1\alpha_2\alpha_3\alpha_4=const >0 \ \
\mbox{\footnotesize are\ tiers\ of\ forest}.
$$
In \cite{vgu1} a stationary equilibriums of this
ecosystem is completely  studied in detail.

Below we examine Nash equilibriums and install them the existence
for a 4-tier mosaic forest ecosystem.

\section{Stationary equilibriums of trier mosaic forest}
\label{S2} \vspace{-4pt}

Stationary equilibriums\ \   $x=x(k,m,a,m)$ of trier mosaic forest
we find by solving the equation
\begin{equation} \label{vgu1}     
\frac{\partial }{\partial x}V(x,k,m,a,w)=0.
\end{equation}
Consider the set
$$
M_V=\{(x,k,m,a,w): \frac{\partial }{\partial
x}V=
$$
$$
=6x^5+4kx^3+3mx^2+2ax+w=0\},
$$
which is contains of maximums, minimums and points of inflection
of function $V_{(k,m,a,w)}(x)=V(x,k,m,a,w)$. All these poins are
stationary equilibriums of given forest ecosystem.

We can change the factors $(k,m,a,w)$ and to get different
stationary equilibriums. In some cases the transition from one
equilibrium to another is  jump $x(k,m,a,w)\to (k',m',a',w')$,
which is called {\it butterfly catastrophe}.

The behavior of the forest ecosystem in such catastrophes is
investigated in \cite{vgu1}.

We shall study behavior of the forest ecosystem under
 the Nash equilibriums.

\section{The algorithm for finding Nash equilibriums}
\label{S2} \vspace{-4pt}

It is natural to consider the differential game with zero sum,
because "winnings" of our players are poorly connected.

If a player forms its control action in the form of only time
function $u(t)$ for the whole duration of the game, then $u(t)$ is
 {\it program control}.  However, the player may select its
control depending on the position $x$ at time $t$ of system.
 In this case, the player constructs a control action
  as a function of $u(t,x)$, which already dependent
  on the position $\{t,x\}$, and for $u(t,x)$ is used the
   term {\it positional  control} \cite{vgu3}.
Often we simply write $u(x)$.

We will look for positional control, positional
 Nash equilibrium.

For differential game with $N$ players
$$
\frac{dx}{dt} = f(x) + \sum_{j = 1}^Ng_j (x) u_j, \ \ \ f (0) = 0,
$$
$$
x \in \R, \ \ u_j \in \R,
$$
$$
J_i(x,u_1,...,u_N) = \int\limits_0^{+\infty} [Q_i(x) + \sum_{j
=1}^N R_{ij}(u_j)^2] dt,
$$
$$
(i = 1,...,N),
$$
$$
Q_i> 0, \ \ R_ {ii}> 0, \ \ R_ {ij} \geq 0,
$$
 the existence problem  of Nash equilibrium
$$
J_i(u_1^*,u_2^*,u_i^*,...,u_N^*) \leq
$$
\begin{equation} \label{nesh2}
J_1(u_1^*,u_2^*,...,u_{i-1}^*,u_i,u_{i+1}^*,...,u_N^*),
\end{equation}  
$$
\forall u_i, \ \ i=1,...,N,
$$
is reduced to extremely complex the problem of finding  positive
definite solutions $V_i(x)>0 $ of nonlinear the  Hamilton-Jacobi
equations
$$ (V_i)'_x(x)f(x) + Q_i(x)-
$$
$$
-\frac{1}{2}(V_i)'_x\sum_{j = 1}^N[G_j(x)]^2 (R_{jj})^{-1}(V_j)'_x
+
$$
\begin{equation} \label {gj}
 + \frac{1}{4}\sum_{j = 1}^N R_{ij}[g_j(x)]^2
[(R_{jj})^{-1}]^2 [(V_j)'_x]^2 = 0,
\end {equation}
Then positional Nash equilibriums are
\begin{equation}
 \label{nesh}  
 u_i^*(x) = u_i(V_i(x)) = -
\frac{1}{2}R_{ii}^{-1} g_i(x)(V_i)'_x.
\end{equation}
$$
(i = 1,...,N).
$$
(see \cite[Theorem 10.4-2]{vgu2}).

\section{Nash equilibrium of the forest ecosystem}
\label{S2} \vspace{-4pt}

In our case $ N = 4 $,  player 1 is  competition of trees  $u_1 =
k$,  player 2 is mosaic factor $u_2 = m$,   player 3 is
anthropogenic interference $u_3 = a$ in the forest ecosystem
(deforestation, fires, and so on.), and, finally,
the player 3 is
soil moisture $ u_4 = w$.

Further
$$
f (x) = -\alpha x^5, \ \ g_1(x) = -4x^3,
$$
$$
g_2(x) = -3x^2 \ \ g_3(x) = -2x, \ \ g_4(x) = -1
$$
and we take
$$
R_{11} = R_{22} = R_{33} = R_{44} = 1, R_{ij} = 0 \ (i \neq j).
$$

The Hamilton-Jacobi equations are:
$$
Q_i + (V_i)'_xf(x) -
$$
\begin{equation} \label {guu1}
 - \frac{1}{2}(V_i)'_x F(x) +
\frac{1}{4}[g_i(x)]^2 [(V_i)'_x]^2 = 0
\end{equation}
$$
(i= 1,2,3,4),
$$
 where
$$
F (x) = \sum_{j = 1}^4 [g_i(x)]^2 (V_i)'_x.
$$

 Assuming that
$$
V_1 (x) = V_2 (x) = V_3 (x) = V_4 (x) = \frac {1}{2} x^2> 0,
$$
we obtain the Hamilton-Jacobi equation in the form
\begin{eqnarray*} \label{guu1}
&Q_1 = \alpha x^6 + 4x^8 + \frac{9}{2} x^6 + 2x^4 + \frac{1}{2} x^2, \\
&Q_2 = \alpha x^6 + 8x^8 + \frac{4}{9} x^6 + 2x^4 + \frac{1}{2} x^2, \\
&Q_3 = \alpha x^6 + 8x^8 + \frac{9}{2} x^6 + x^4 + \frac {1}{2} x^2,\ \  \\
&Q_4 = \alpha x^6 + 8x^8 + \frac{9}{2} x^6 + 2x^4 +\frac{1}{4}x^2.
\end{eqnarray*}
Since all functions $Q_i$ are positive definite, then the
Hamilton-Jacobi equations are held for those features and for
functions $V_i$ that were selected above.

 Therefore by Theorem 10.4-2 of \cite {vgu2} we have a Nash
equilibrium
\begin{equation} \label{neshh}
\begin{split}
k^* = 2x^4, \ \ \ m^* = \frac{3}{2}x^3, \\
a^* = x^2, \ \ \ w^* = \frac{1}{2}x,\ \
\end{split}
\end{equation}
found by the formulas (\ref{nesh}).

 We have the following winning / losing functions:
\begin{eqnarray*} \label{nns}
J_1(x,k,m,a,w) = \int\limits_0^{+\infty}[Q_1(x) + k^2]dt, \\
J_2(x,k,m,a,w) = \int\limits_0^{+\infty}[Q_2(x) + m^2]dt, \\
J_3(x,k,m,a,w) = \int\limits_0^{+\infty}[Q_3(x) + a^2]dt, \\
J_4(x,k,m,a,w) = \int\limits_0^{+\infty}[Q_4(x) + w^2]dt,
\end{eqnarray*}

Productivity $x$ under the Nash equilibrium (\ref{neshh}) is found
by integration of equation (\ref{vga7})--(\ref{vga8}) and is
satisfies the equatution
$$
\int\frac{x^{-1}dx}{8x^6+(9/2+\alpha)x^4+2x^2+1/2}=
$$
\begin{equation}\label{vgu9}
    =-t+C,
\end{equation}
where $C$ is constant of integration, or
$$
2\ln(x)-
$$
$$
-\sum_R\frac{(16R^2+9R+2Ra+4)\ln(x^2-R)}{(48R^2+18R+4Ra+4)}=
$$
$$
= -t+C,
$$
where
$$
R\ \ \mbox{is root of}\ \  16Z^3+(9+2a)Z^2+4Z+1=0.
$$
For $\alpha=0,0007$, i.~e. for forest with 70\% of  upper tier
mass of and for 10\% in third others we have the following
solution:
$$
2.0\ln(x)-0.772183\ln(x^2+0.354611)-
$$
$$
-0.113908\ln((x^2+0.103988)^2+ 0.165435)+
$$
$$
+0.731466\arctan(\frac{0.406737}{x^2+0.103988})=
$$
$$
= -t+C.
$$
In the case $C=-100$ this solution is presented at fig.~1.

\begin{figure}[h!]\label{les1}
\centering\includegraphics[height=4cm]{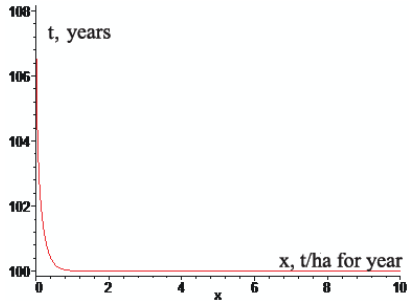} \vspace{-0.2cm}

\medskip
{\footnotesize \caption{Forest productivity in the condition of
the Nash equilibrium (\ref{neshh}).} }
\end{figure}
We see that over time productivity of phytocenosis falls. In other
words, the forest is degraded. But degradation  comes no sooner
than 100 years. We have a climax forest.

\section{Conclusion}

We have shown that it is possible to apply the theory of
differential games to the study of forest ecosystems. We have
shown that in such ecosystem there exist the Nash equilibriums
that is installed in the system when reached some defined mediated
the connection between the external factors affecting  on the
productivity of the forest.

As further research is necessary and useful to determine which
forests and in some cases are in Nash equilibrium, and how it is
expressed in terms of the traditional  science on forests and
forest ecosystems.

\vspace{10pt} \noindent {\bf Acknowledgements:} \ The authors
thank professor N.A.~Kalinenko for support of our research.

\end{document}